\documentclass[fleqn,twoside]{article}
\usepackage{amsmath}
\usepackage{subfigure}
\usepackage[headings]{espcrc2}

\readRCS
$Id: espcrc2.tex,v 1.2 2004/02/24 11:22:11 spepping Exp $
\usepackage{graphicx}
\usepackage[figuresright]{rotating}

\newcommand{\tanb}{\mathrm{tan}\beta}
\newcommand{\gev}{\mathrm{GeV}}
\newcommand{\Mhmax}{M_{h,{\mathrm{max}}}}
\newcommand{\B}{\ensuremath{\mathbf}}

\title{
  \quad\vskip-3.3cm \hfill {\normalsize
  \begin{tabular}[t]{l}
                     \rule{0ex}{1ex}MAN/HEP/2009/5
  \end{tabular}}
  \vskip1.5cm 
Central Exclusive Production at the LHC}

\author{J.R. Forshaw\address[UoM]{School of Physics \&
    Astronomy, University of Manchester, Oxford Road, Manchester M13
    9PL, UK.}\thanks{Talk presented at the workshop ``New Trends in
    HERA Physics'', Ringberg Castle, Tegernsee, 5--10 October 2008.}}

\runtitle{Central Exclusive Production}
\runauthor{J.R Forshaw}

\begin{document}

\begin{abstract}
After a brief resum\'e of the theory underpinning the central
exclusive process (CEP) $pp \to
p+H+p$, attention is focussed upon Higgs bosons produced in the
Standard Model, the MSSM and the NMSSM. In all cases, CEP adds
significantly to the physics potential of the LHC and in some
scenarios it may be crucial.   
\vspace{1pc}
\end{abstract}

\maketitle

\setcounter{footnote}{0}
\section{Introduction}
\begin{figure}[htb]
\centering
\includegraphics[width=9.5pc]{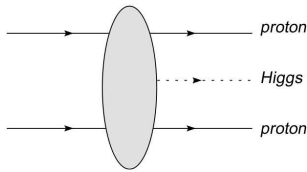}
\caption{Central Exclusive Production of a Higgs boson.}
\label{fig:cep}
\end{figure}

The idea to install detectors far from the interaction point at
CMS and/or ATLAS with the capacity to detect protons scattered through
small angles has gained a great deal of attention in recent
years and the report presented in \cite{Albrow:2008pn} constitutes a
significant milestone on the road to CEP physics at the LHC. In this talk, I
should like to focus attention in particular on Higgs boson
production, as illustrated in Fig.~\ref{fig:cep}. For a much more
extensive survey of the physics that can be studied after installing
forward detectors see \cite{Albrow:2008pn}.
Detection of both protons by detectors located 420m from the IP at
ATLAS/CMS has the virtue that central systems with masses up to
200~GeV can be observed with an event-by-event precision of around 2
to 3 GeV, and adding also 220m detectors extends the reach to much
higher masses. The clean environment of CEP generally makes for
reduced backgrounds (even in the presence of significant amounts of
pile-up) and that is often aided by the fact that the centrally
produced system is predominantly in a $J_z=0$, C-even, P-even
state. Of course having such a spin-parity filter also provides an
excellent handle on the nature of any new physics. For very little
extra cost, forward detectors promise to significantly enhance the
physics potential of the LHC.

That said, CEP is not without its challenges. The theory is difficult,
triggering can be tricky, signal rates for new physics are often low,
new detectors need building and installing, and pile-up needs to be
brought under control. With the arrival of data on CEP from CDF at the
Tevatron, confidence is building in the theoretical modelling and
extensive studies have demonstrated that all of the other challenges
can be met, e.g. see \cite{Albrow:2008pn}.

\begin{figure}[htb]
\centering
\includegraphics[width=9.5pc]{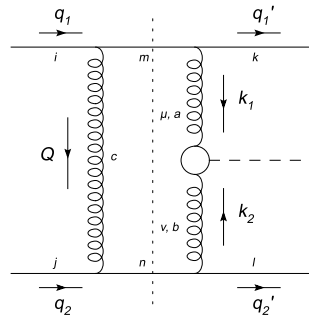}
\caption{The relevant lowest order Feynman diagram for $qq \to q+H+q$.}
\label{fig:qHq}
\end{figure}

Without further ado, let us very quickly review the Durham model for
CEP, more details can be found in \cite{Khoze:2001xm,Forshaw:2005qp}. The calculation starts from
the easier to compute parton level process $qq \to q+H+q$ shown in
Fig.~\ref{fig:qHq}. The Higgs is produced via a top quark loop and a
minimum of two gluons need to be exchanged in order that no colour be transferred between the
incoming and outgoing quarks. The real part of the amplitude is small
and the imaginary part can be determined by considering only the cut
diagram in Fig.~\ref{fig:qHq}. The calculation of the amplitude is straightforward:
\begin{eqnarray}
{\mathrm{Im}} 
A^{ik}_{jl} &=& \int d(PS)_2 \; \delta((q_1-Q)^2) \delta((q_2+Q)^2)
\nonumber \\ & & \hspace*{-2cm}
\frac{2 g q_1^{\alpha} \; 2 g q_{2 \alpha}}{Q^2} \; \frac{2g q_1^{\mu}}{k_1^2} 
\; \frac{2g q_2^{\nu}}{k_2^2} \; V^{ab}_{\mu \nu} \; 
\tau^c_{im} \tau^c_{jn} \tau^a_{mk} \tau^b_{nl}~.
\end{eqnarray}
We write $Q = \alpha q_1 + \beta q_2 + Q_T$. The delta functions fix
the cut quark lines to be on-shell, which means that $\alpha \approx
-\beta \approx \B{Q}_T^2/s \ll 1$ and $Q^2 \approx Q_T^2 \equiv -
\B{Q}_T^2$. As always, we neglect terms that are energy suppressed such as the product
$\alpha \beta$. In the Standard Model, the Higgs production vertex is
\begin{equation}
V^{ab}_{\mu \nu} = \delta^{ab} \left( g_{\mu \nu} - \frac{k_{2 \mu} k_{1 \nu}}{k_1 \cdot k_2}
\right) V~,
\end{equation}
where $V = m_H^2 \alpha_s/(4 \pi v) F(m_H^2/m_t^2)$ and $F \approx 2/3$ provided the Higgs is
not too heavy. The Durham group also include a NLO K-factor correction to this vertex.

We can compute the contraction $q_1^{\mu} V^{ab}_{\mu \nu} q_2^{\nu}$ either directly or
by utilising gauge invariance, which requires that $k_1^{\mu} V^{ab}_{\mu \nu} = 
k_2^{\nu} V^{ab}_{\mu \nu} = 0$. Writing\footnote{We can do this because $x_i 
\sim m_H/\surd{s}$ whilst the other Sudakov components are $\sim Q_T^2/s$.} 
$k_i = x_i q_i + k_{i T}$ yields
\begin{equation}
q_1^{\mu} V^{ab}_{\mu \nu} q_2^{\nu} \approx 
\frac{k_{1T}^{\mu}}{x_1} \frac{k_{2T}^{\nu}}{x_2} V^{ab}_{\mu \nu} \approx 
\frac{s}{m_H^2} k_{1T}^{\mu} k_{2T}^{\nu} V^{ab}_{\mu \nu} 
\end{equation}
since $2 k_1 \cdot k_2 \approx x_1 x_2 s \approx m_H^2$. Note that it is
as if the gluons which fuse to produce the Higgs are transversely polarized,
$\epsilon_i \sim k_{iT}$. Moreover,
in the limiting case that the outgoing quarks carry no transverse momentum
$Q_T = -k_{1T} = k_{2T}$ and so $\epsilon_1 = -\epsilon_2$. This is
an important result; it generalizes to the statement that the centrally produced
system should have a vanishing $z$-component of angular momentum in the limit that
the protons scatter through zero angle (i.e. $~ q_{iT}'^{2} \ll
Q_T^2$). Since we are interested in  very small angle
scattering this selection rule is effective. One immediate consequence
is that the Higgs decay to $b$-quarks may now be viable. This is because, for massless quarks, 
the lowest order $q \bar{q}$ background vanishes identically (it does not vanish at NLO). The
leading order $b \bar{b}$ background is therefore suppressed by a factor $\sim m_b^2/m_H^2$.

Returning to the task in hand, we can write ($y$ is the rapidity of
the Higgs)
\begin{eqnarray}
\frac{d\sigma}{d^2 \B{q}_{1T}' d^2 \B{q}_{2T}' dy} &\approx&
\left(\frac{N_c^2-1}{N_c^2}\right)^2 \frac{\alpha_s^6}{(2 \pi)^5}
\frac{G_F}{\surd{2}}
\nonumber \\ && \hspace*{-2cm} \times
\left[ \int \frac{d^2 \B{Q}_T}{2 \pi} 
\frac{\B{k}_{1T} \cdot \B{k}_{2T}}{\B{Q}_T^2 \B{k}_{1T}^2  \B{k}_{2T}^2} 
\frac{2}{3} \right]^2~. \label{eq:qHq}
\end{eqnarray}
We are mainly interested in the
forward scattering limit whence 
$$ 
\frac{\B{k}_{1T} \cdot \B{k}_{2T}}{\B{Q}_T^2 \B{k}_{1T}^2  \B{k}_{2T}^2} 
\approx -\frac{1}{\B{Q}_T^4}.
$$
As it stands, the integral over $Q_T$ diverges. Let us not worry about that for now and
instead turn our attention to how to convert this parton level cross-section into the
hadron level cross-section we need.

\begin{figure}[htb]
\centering
\includegraphics[width=12pc]{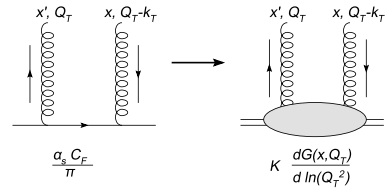}
\caption{The recipe for replacing the quark line (left) by a proton line (right).}
\label{fig:updf}
\end{figure}

What we really want is the hadronic matrix element which
represents the coupling of two gluons into a proton, and this is really an off-diagonal
parton distribution function. At present we don't have much knowledge of these
distributions, however we do know the diagonal gluon distribution function.  
Fig.~\ref{fig:updf} illustrates the Durham prescription for coupling the two gluons
into a proton rather than a quark. The factor $K$ would equal unity if $x'=x$ and
$k_T=0$, which is the diagonal limit. That we should, in the amplitude, replace a factor of 
$\alpha_s C_F/\pi$ by $\partial G(x,Q_T)/\partial \ln  Q_T^2$ can be easily derived
starting from the DGLAP equation for evolution off an initial quark distribution given by
$q(x) = \delta(1-x)$. The Durham approach makes use of a result derived in \cite{Shuvaev:1999ce}
which states that in the case $x' \ll x$ and $k_T^2 \ll Q_T^2$ the off-diagonality can
be approximated by a multiplicative factor, $K$. Assuming a Gaussian form factor suppression
for the $k_T$-dependence they estimate that
\begin{equation}
K \approx e^{-b k_T^2/2} \frac{2^{2\lambda+3}}{\surd{\pi}} 
\frac{\Gamma(\lambda+5/2)}{\Gamma(\lambda+4)} \label{eq:offdiag}
\end{equation}
and this result is obtained assuming a simple power-law behaviour of the gluon density,
i.e. $G(x,Q) \sim x^{-\lambda}$. For the production of a 120 GeV Higgs boson at the 
LHC, $K \sim 1.2 \times e^{-b k_T^2/2}$. In the cross-section, the off-diagonality therefore
provides an enhancement of $(1.2)^4 \approx 2$. Clearly the current lack of knowledge of the
off-diagonal gluon is one source of uncertainty in the
calculation. The slope parameter $b$ is fixed by assuming the same
$k_T$-dependence as for diffractive $J/\psi$ production\footnote{It
  turns out that the typical $Q_T \sim  1.5$ GeV for a 120 GeV
  Higgs.}, i.e. $b \approx 4$~GeV$^{-2}$.

Thus, after integrating over the transverse momenta of the scattered protons we have
\begin{eqnarray}
\frac{d \sigma}{dy} &\approx& \frac{1}{256 \pi b^2} \frac{\alpha_s G_F
  \surd{2}}{9}
\nonumber \\ &&  \times
\left[ \int \frac{d^2 \B{Q}_T}{\B{Q}_T^4} \; f(x_1,Q_T) f(x_2,Q_T) \right]^2
\end{eqnarray}
where $f(x,Q) \equiv \partial G(x,Q)/\partial \ln Q^2$ and we have neglected
the exchanged transverse momentum in the integrand. 

Now it is time to worry about the fact that our integral diverges in the infra-red.
Fortunately I have missed some crucial physics. The lowest order diagram is not
enough, virtual graphs possess logarithms in the ratio $Q_T/m_H$ which are very
important as $Q_T \to 0$; these logarithms need to be summed to all orders. This is 
Sudakov physics: thinking in terms of real emissions we must be sure to forbid
real emissions into the final state. Let's worry about real gluon emission off
the two gluons which fuse to make the Higgs. The emission probability for a
single gluon is (assuming for the moment a fixed coupling $\alpha_s$)
\begin{eqnarray}
\frac{C_A \alpha_s}{\pi} \int_{Q_T^2}^{m_H^2/4} \frac{dp_T^2}{p_T^2}
\int_{p_T}^{m_H/2} \frac{dE}{E} && \nonumber \\ && \hspace*{-2cm} = \frac{C_A \alpha_s}{4\pi} \ln^2 
\left( \frac{m_H^2}{Q_T^2} \right).
\end{eqnarray}
The integration limits are kinematic except for the lower limit on the $p_T$ integral.
The fact that emissions below $Q_T$ are forbidden arises because the gluon not
involved in producing the Higgs completely screens the colour charge of the fusing
gluons if the wavelength of the emitted radiation is long enough, i.e. if $p_T < Q_T$.
Now we see how this helps us solve our infra-red problem: as $Q_T \to 0$ so the
screening gluon fails to screen and real emission off the fusing gluons cannot be
suppressed. To see this argument through to its conclusion we realise that multiple
real emissions exponentiate and so we can write the non-emission probability as
\begin{equation}
e^{-S} = 
\exp \left( -\frac{C_A \alpha_s}{\pi} \int_{Q_T^2}^{m_H^2/4} \frac{dp_T^2}{p_T^2}
\int_{p_T}^{m_H/2} \frac{dE}{E} \right). \label{eq:Sudakov1}
\end{equation}
As $Q_T \to 0$ the exponent diverges and the non-emission probability vanishes faster than
any power of $Q_T$. In this way our integral over $Q_T$ becomes
\begin{equation}
\int \frac{dQ_T^2}{Q_T^4} f(x_1,Q_T) f(x_2,Q_T) \; e^{-S}~, \label{eq:DLLA}
\end{equation}
which is finite. 

Now Eq.~(\ref{eq:Sudakov1}) is correct only so far as the leading double logarithms. It is
of considerable practical importance to correctly include also the single logarithms. 
To do this we must re-instate the running of $\alpha_s$ and allow for the possibility
that quarks can be emitted. Including this physics means we ought to use
\begin{eqnarray}
e^{-S} &=&  \exp \left( -\int_{Q_T^2}^{m_H^2/4} \frac{dp_T^2}{p_T^2} 
\frac{\alpha_s(p_T^2)}{2 \pi} \right.  \\ && \left. \times \int^{1-\Delta}_{0} dz \; [
z P_{gg}(z) + \sum_q P_{qg}(z) ] \nonumber
\right) \label{eq:Sudakov2}
\end{eqnarray}
where $\Delta = 2 p_T/m_H$, and $P_{gg}(z)$ and $P_{qg}(z)$ are the 
leading order DGLAP splitting functions. To correctly sum all single logarithms
requires some care in that what we want is the distribution of gluons in $Q_T$
with no emission up to $m_H$, and this is in fact \cite{Martin:2001ms}
$$
\tilde{f}(x,Q_T) = \frac{\partial}{\partial \ln Q_T^2} \left( e^{-S/2} \; G(x,Q_T) \right).
$$
The integral over $Q_T$ is therefore
$$
\int \frac{dQ_T^2}{Q_T^4} \tilde{f}(x_1,Q_T)
\tilde{f}(x_2,Q_T)~, \label{eq:Qt}
$$
which reduces to Eq.~(\ref{eq:DLLA}) in the double logarithmic
approximation.

Before we can go ahead and compute the cross-section we need to
introduce the idea of gap survival. The Sudakov factor has allowed us
to ensure that the exclusive nature of the final state is not spoilt
by perturbative emission off the hard process. What about
non-perturbative particle production? The protons can in principle
interact quite  apart from the perturbative process discussed hitherto and this interaction 
could well lead to the production
of additional particles. We need to account for the probability that such emission
does not occur. Provided the hard process leading to the production of the
Higgs occurs on a short enough timescale, we might suppose that the physics which
generates extra particle production factorizes and that its effect can be accounted
for via an overall factor multiplying the cross-section we have just calculated. This
is the ``gap survival factor''.  The gap survival, $S^2$, is thus defined by
$$ d\sigma(p+H+p|\mathrm{no~soft~emission}) = d\sigma(p+H+p) \times S^2 $$
where $d\sigma(p+H+p)$ is the differential cross-section computed above. The task is to
estimate $S^2$. Clearly this is not straightforward since we cannot utilize QCD
perturbation theory. That said, data on a variety
of processes observed at HERA, the Tevatron and the LHC can help us improve our 
understanding of ``gap survival'' and to date the HERA and Tevatron
data do support the idea of gap survival. For the purposes of this
talk we will presume to know the gap survival factor and that $S^2 = 3\%$ for CEP
at the LHC, e.g. see \cite{Martin:2008nx} for an overview.

\begin{figure}
\begin{center}
\includegraphics[width=0.4\textwidth]
                {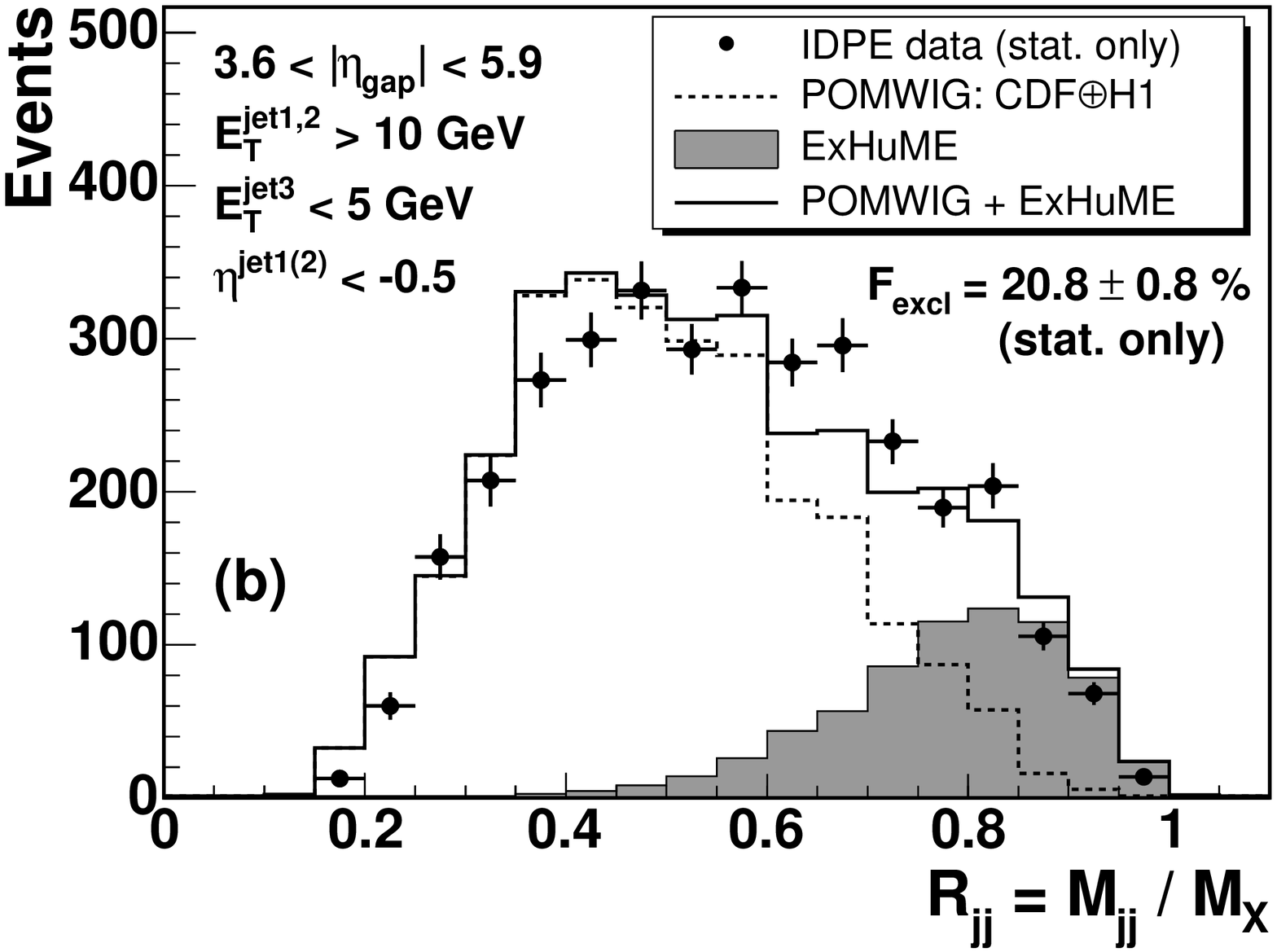}
\includegraphics[width=0.4\textwidth]
                {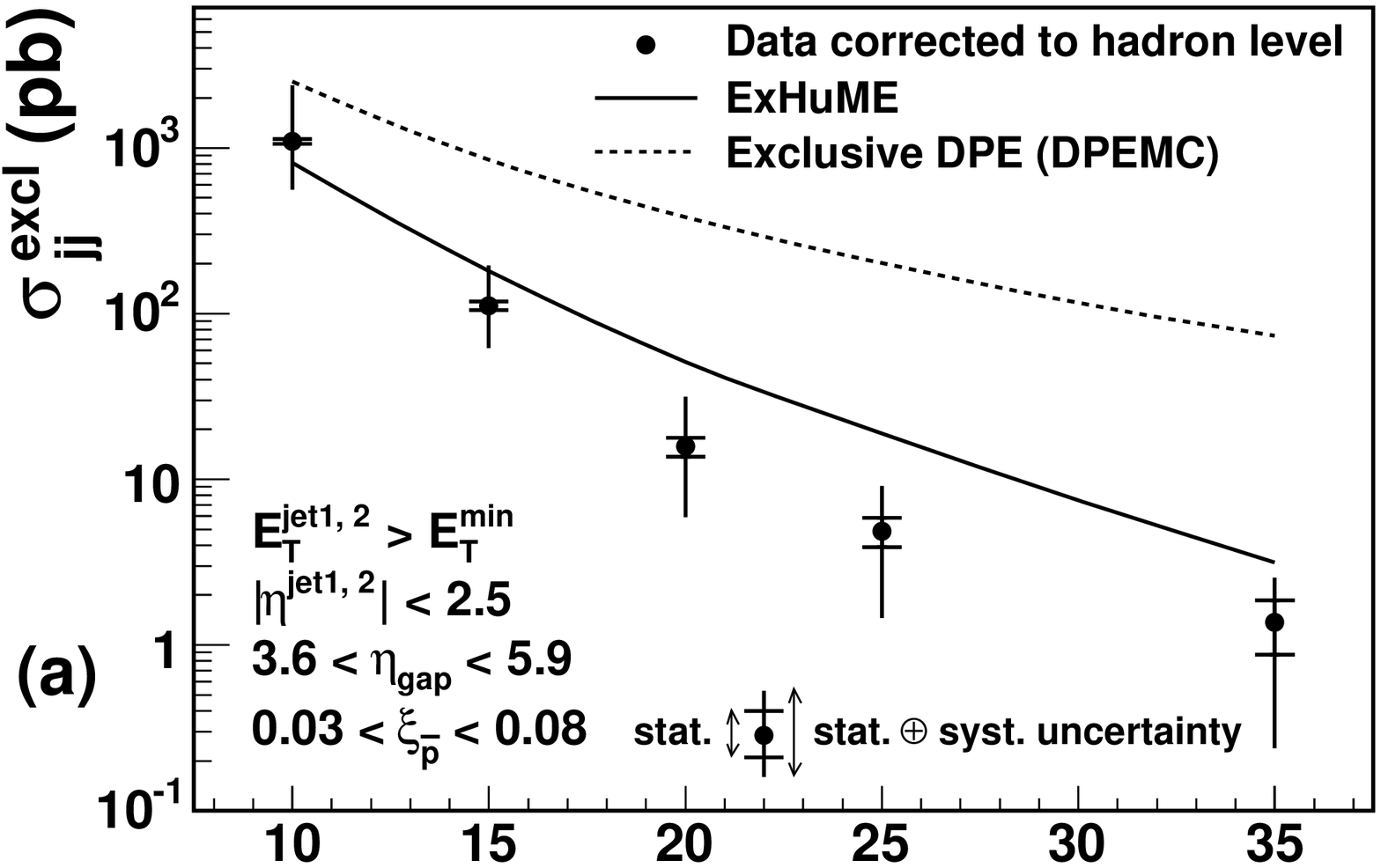}
\caption{CDF data on CEP dijet production. Figures from \cite{Aaltonen:2007hs}.}
\label{fig:CDF}
\end{center}
\end{figure}

Recently, the CDF collaboration has observed a $6\sigma$ excess of CEP of dijets at the
Tevatron \cite{Aaltonen:2007hs}. The agreement with the theory (as implemented in the ExHuME
monte carlo generator \cite{Monk:2005ji}) is very good, as illustrated
in Fig.~\ref{fig:CDF}. CDF also sees a suppression of quark jets in
the exclusive region (high $R_{jj}$), in accord with theoretical expectations.

\section{Higgs: SM and MSSM}
\begin{figure}[htb]
\centering
\includegraphics[width=0.4\textwidth]{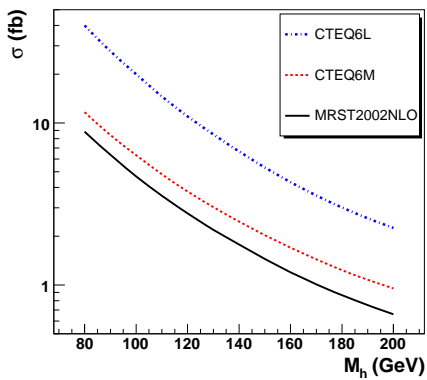}
\caption{CEP cross-section for Standard Model Higgs producton. Figure
  from \cite{Albrow:2008pn}.}
\label{fig:SM}
\end{figure}

Fig.~\ref{fig:SM} shows how the cross-section for producing a SM Higgs
boson varies with Higgs mass (and for different gluon distribution
functions). The cross-section is small and leads to low production
rates. That said, a SM Higgs with mass above 120~GeV should be observable in the $WW^*$
channel with 300~fb$^{-1}$ of data (which is around 3 years of high
luminosity running) \cite{Albrow:2008pn,Cox:2005if}. The gold-plated
fully leptonic channel has very low backgrounds and has the advantage
that one can still use the forward detectors to measure the mass. 

The $b\bar{b}$ channel is much more
challenging. Triggering in this case would certainly benefit from
having 220m detectors in place but even then one relies on optimistic
scenarios for the production cross-section, detector acceptance and
trigger efficiency. Nevertheless, it ought to be born in mind that CEP
may be the only way to explore this channel at the LHC. 

\begin{figure}
\begin{center}
\includegraphics[width=0.45\textwidth]
                {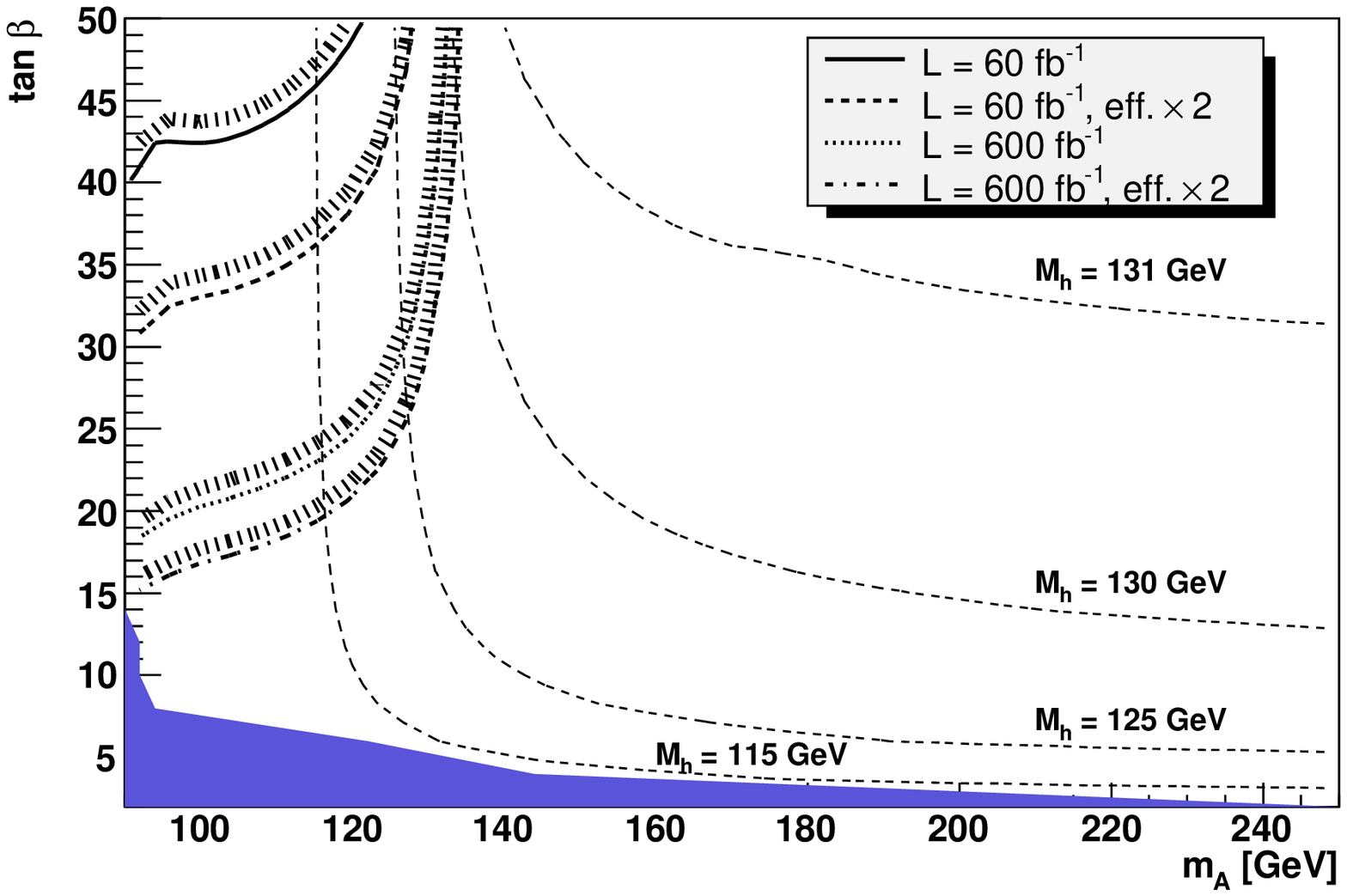}
\includegraphics[width=0.45\textwidth]
                {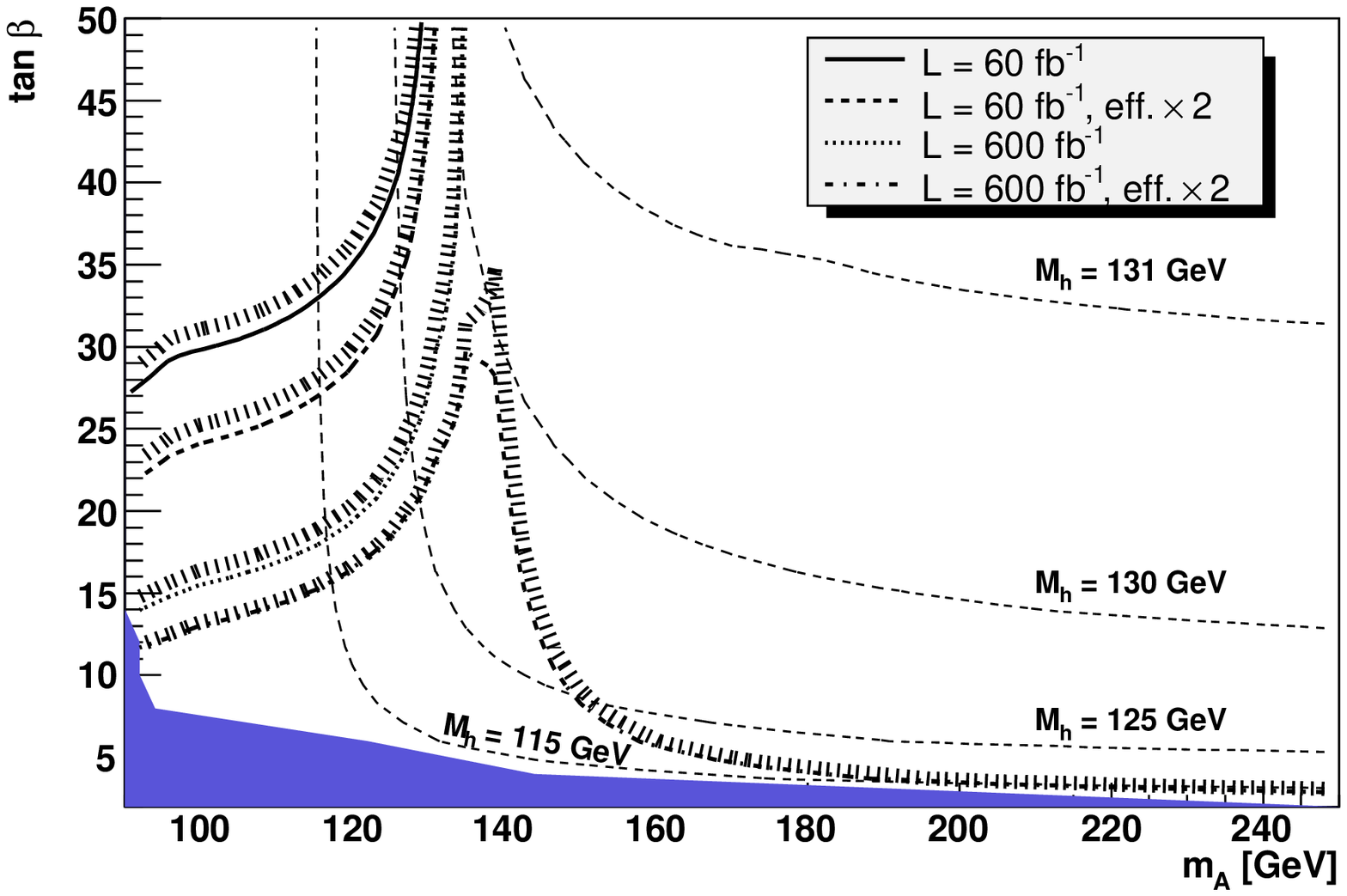}
\caption{
$5 \sigma$ discovery contours (upper plot) and contours of $3 \sigma$
statistical significance (lower plot) for the $h \to b \bar b$ channel in
CEP in the $M_A$ - \rm{tan}$\beta$ plane of the MSSM within the $\Mhmax$
benchmark scenario for different luminosity scenarios as described in the text~\cite{Heinemeyer:2007tu}.  
The values of the mass of the light CP-even Higgs boson, $M_h$, are
indicated by contour lines. No pile-up background assumed. 
The dark shaded (blue) region corresponds to the parameter region that
is excluded by the LEP Higgs boson
searches~\cite{Barate:2003sz,Schael:2006cr}. Figure from \cite{Albrow:2008pn}.
}
\label{fig:heinemeyer2}
\end{center}
\end{figure}

\begin{figure}
\centering
\includegraphics[width=.45\textwidth]{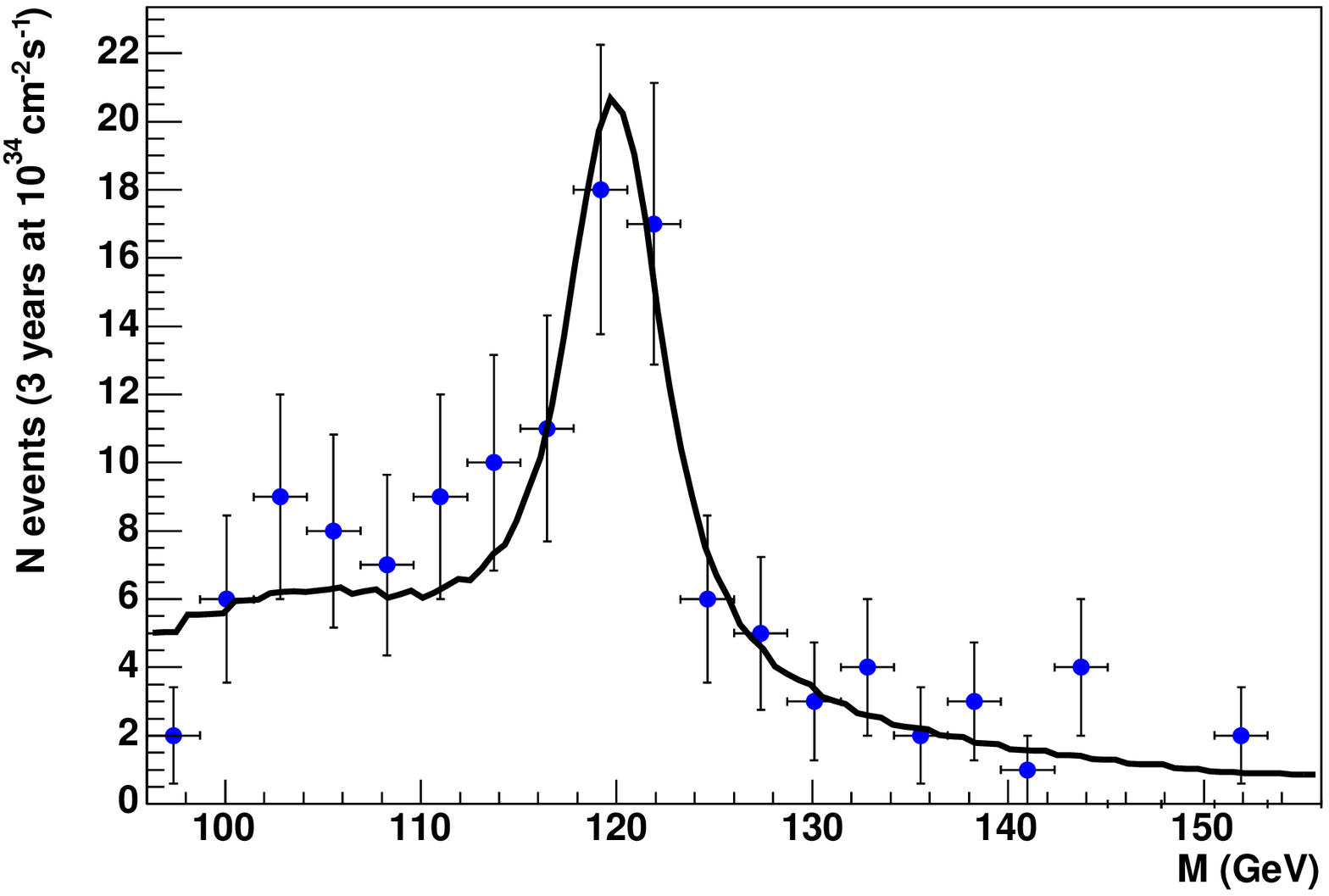}\\
\includegraphics[width=.45\textwidth]{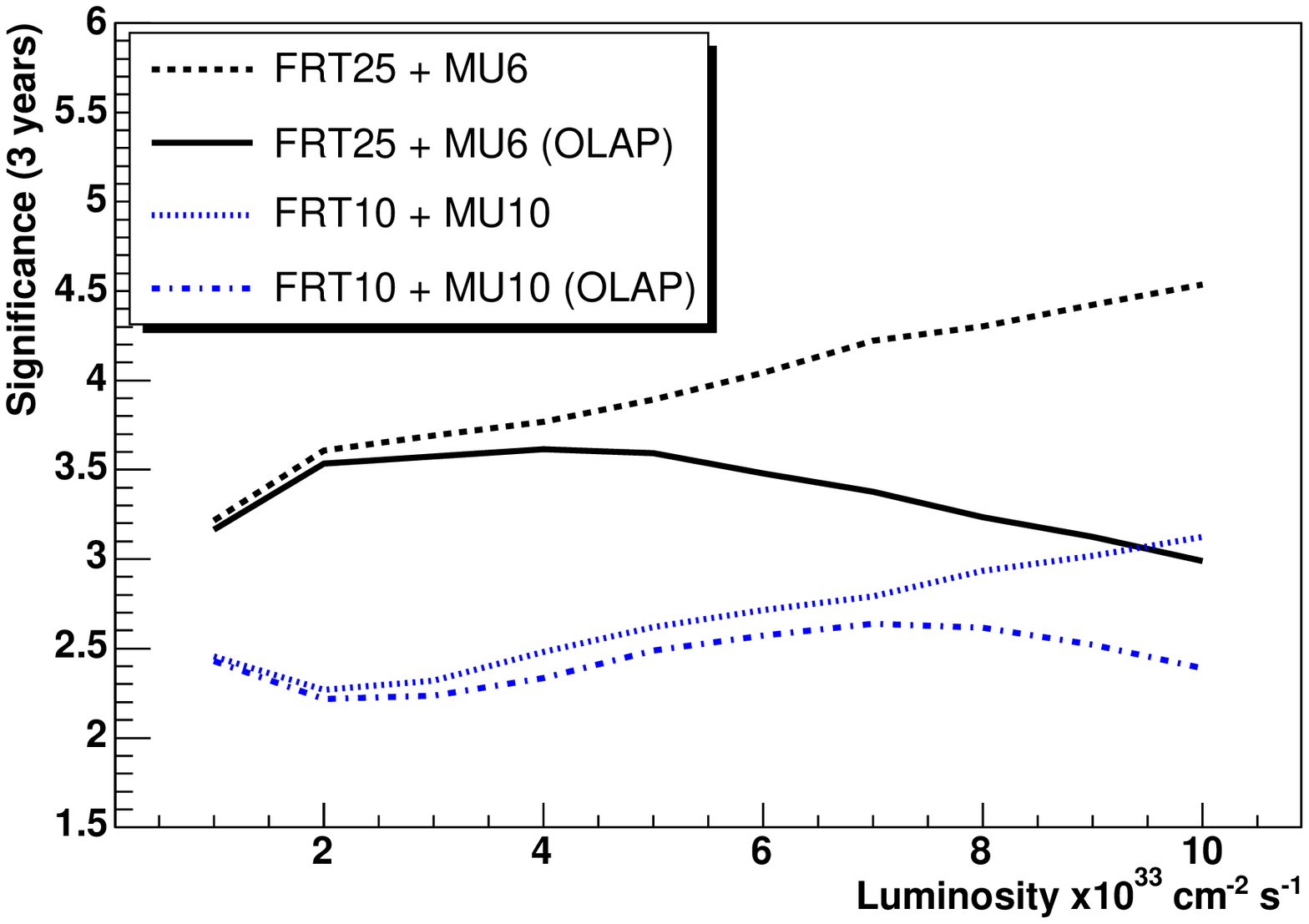}
\caption{Upper: Typical mass fit for the 120~GeV MSSM $h\rightarrow b\bar{b}$ 
for 3 years of data taking at $10^{34}~$cm$^{-2}~$s$^{-1}$ after removing the overlap 
background contribution completely with improved timing detectors. The significance is 
$5 \sigma$ for these data. Lower: Significance of the measurement of the 120 GeV MSSM 
Higgs boson versus luminosity, for two different combinations of muon
(MU6, MU10) and fixed-jet-rate (FRT25, FRT10) triggers and with an improved (baseline) FP420 
timing design (OLAP labels). Figure from \cite{Albrow:2008pn}.}
\label{fig:peak2}
\end{figure}

In contrast, the $b\bar{b}$ channel becomes much more exciting in
certain MSSM scenarios. Rates are strongly enhanced at large $\tan
\beta$ and small $m_A$, and the potential to measure the $hbb$ Yukawa
coupling is a strong selling point for CEP and a pre-requisite to
determining any Higgs-boson coupling at the LHC (rather than just ratios of
couplings). Fig.~\ref{fig:heinemeyer2} shows the region of parameter space\footnote{In the $M_h^{\mathrm{max}}$ scenario with $\mu = +200$~GeV.}
in which one could observe $h \to b\bar{b}$ using CEP
\cite{Heinemeyer:2007tu} with different amounts of integrated
luminosity. A similar pair of plots can be produced for $H \to b \bar{b}$, see \cite{Heinemeyer:2007tu}. Fig.~\ref{fig:peak2} shows the result of an in-depth analysis of one
particular point in the $m_A-\tan\beta$ plane ($\tan\beta =40$ and
$m_A=120$~GeV) \cite{Cox:2007sw}. The details of the two analyses can
be found in \cite{Albrow:2008pn,Heinemeyer:2007tu,Cox:2007sw} but the key point is that they are in general agreement

The curves in Fig.~\ref{fig:peak2} correspond to different trigger
scenarios. They also indicate the influence of pile-up and, in particular,
the overlap background (OLAP), in which the signal is faked by a
coincidence of events, one that produces the central system (which
fakes the Higgs decay) and one or more diffractive events that are
able to produce protons in the forward detectors, e.g. a
three-fold co-incidence of two single diffractive events with a $pp
\to b\bar{b}X$ event. Use of fast-timing detectors allows a
significant reduction in the OLAP background, as the primary vertex
can be pinpointed to high accuracy. Improvements in the fast-timing
could potentially eliminate the OLAP background completely and allow a
$5\sigma$ discovery with 3 years of high luminosity data taking (the
mass peak is illustrated in the upper plot in Fig.~\ref{fig:peak2}).  

\section{Higgs: NMSSM}
To conclude, I would like to take a slightly more in-depth look at the
possibilities for CEP of NMSSM Higgs bosons. More details can be found
in  \cite{Forshaw:2007ra}. The NMSSM is an extension of the MSSM that
solves the $\mu$-problem, and also the little hierarchy problem, by
adding a gauge-singlet superfield $\hat{S}$ to the MSSM such that the $\mu$
term is now dynamical in origin, arising when the scalar member of
$\hat{S}$ aquires a vev. The $\mu$ problem is solved since $\mu$ is no
longer fundamental and therefore no longer naturally of order the GUT
scale (as is the case if it is the only dimensionful parameter in the
superpotential). The little heirarchy problem is also solved because a
lighter Higgs is allowed, thereby taking the pressure off the stop
mass. More specifically, the lightest scalar Higgs can decay
predominantly to two pseudo-scalar Higgses and the branching ratio to
$b$-quarks is correspondingly suppressed, thereby evading the 114 GeV
bound from LEP\footnote{It drops to 86 GeV.}. Having a lighter Higgs
means that the stop mass does not need to be so large, and that is preferred
given the value of $M_Z$. 

\begin{figure}[h]
\begin{center}
\includegraphics[width=0.33\textwidth,angle=90]{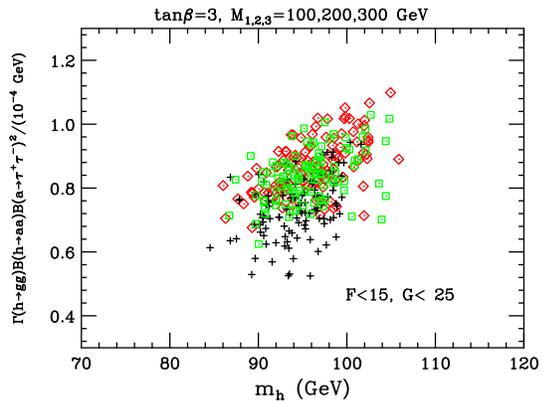}
\caption{$\Gamma_{\mathrm{eff}}$ in units of $10^{-4}\gev^2$ versus $m_h$ for
  $\tanb=3$. Point coding: (red) diamonds =
  $2m_\tau<m_a<7.5\gev$; (green) squares = $7.5\gev\leq m_a <8.8\gev$;
  (black) pluses = $8.8\gev\leq m_a<2m_b$. Figure from \cite{Forshaw:2007ra}.}
\label{fig:scatter}
\end{center}
\end{figure}

The Higgs sector of the NMSSM extends that
of the MSSM by adding an extra pseudo-scalar Higgs and an extra scalar
Higgs: crucially $\hat{S}$ is a gauge singlet and hence $h
\to aa$ can dominate with a light $a$ (i.e. below the threshold
for $a \to b\bar{b}$).
\begin{figure}[h]
\begin{center}
\includegraphics[width=0.36\textwidth]{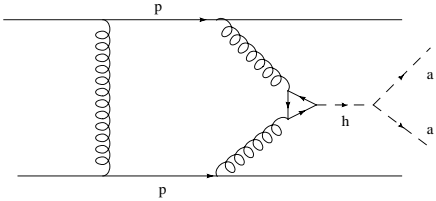}
\caption{CEP of an NMSSM Higgs.}
\label{fig:nmssm}
\end{center}
\end{figure}
Freed of the heavy stop, it is most natural to have a light Higgs with
a reducing branching ratio to $b$-quarks, as illustrated in
Fig.~\ref{fig:scatter}. $F$ and $G$ are measures of fine-tuning, 
so the points in this scatter plot are supposed to
represent most natural scenarios in the NMSSM. Our attention will
focus on one such point, with $m_h = 93$~GeV and $m_a = 9.7$~GeV with
BR$(h \to aa)=92\%$ and BR$(a\to \tau \tau)=81\%$ \cite{Ellwanger:2005dv}. The lightness of
the pseudo-scalar $a$ means that the $h$ decays predominantly to four
taus. Should such a decay mode be dominant at the LHC, standard
search strategies would fail and, as we shall see, CEP (as illustrated
in Fig.~\ref{fig:nmssm}) could provide the discovery channel.
This ``natural'' scenario of the NMSSM
has two additional bonus features that one might draw attention to:
1. a light Higgs is preferred by the precision electroweak data
(recall the best fit value is somewhat below 100 GeV); 2. a 100 GeV
Higgs with a reduced (10\%) branching ratio to $b$-quarks naturally
accommodates the existing $2.3\sigma$ LEP excess in $e^+e^- \to
Zb\bar{b}$ \cite{Dermisek:2005ar,Dermisek:2005gg}.
 
To detect the four-tau decay of an NMSSM Higgs using CEP, we need
first to trigger the event and to that end demand that at least one of
the taus decays to produce a sufficiently high $p_T$ muon. The muon
then defines a vertex which can be used, in conjunction with the
(picosecond) fast timing of the 420m detectors, to reject pile-up
related backgrounds. The detailed analysis is outlined in
\cite{Forshaw:2007ra}, here we shall just highlight the key
features. Table 1 shows how the signal (CEP) and backgrounds (DPE,
OLAP and QED) are affected by the cuts imposed. The top line of the
table is the cross-section after imposing that there be at least one muon with
$p_T > 6$~GeV, which is the nominal minimum value to trigger at level 1 in
ATLAS\footnote{It will turn out that a higher cut of 10 GeV is
  preferred in the subsequent analysis.} 
and the condition that both protons be detected in the
420m detectors. There is also a loose cut on the invariant mass of the
central system. Of the remaining cuts, I would like to single out the
``$N_{\mathrm{ch}}=4$ or 6'' cut. 
\begin{figure}[t]
\centering
\includegraphics[width=0.45\textwidth]{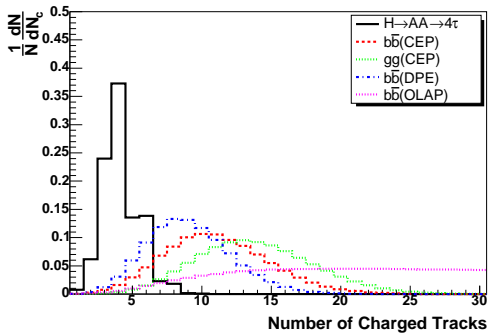}
\caption{The expected number of charged tracks reconstructed by the ATLAS inner detector with $p_T>6$~GeV and $|\eta|<2.5$.}
\label{fig:nc46}
\end{figure}
The charged track ($N_{\mathrm{ch}}$) cut is noteworthy because it can
be implemented at the highest LHC luminosities: we cut on exactly 4 or
6 charged tracks that point back to the vertex defined by the
muon. Pile-up events do add extra tracks (to both the signal and background), but they do not often
coincide with the primary vertex (i.e. within a 2.5~mm window) and do
not spoil the effectiveness of this cut. The number of charged tracks
in signal and background events is illustrated in
Fig.~\ref{fig:nc46}. The ability to make such hard cuts on charged tracks could be of a much
wider utility than this analysis (e.g. in defining a jet veto in Higgs
plus two jets production). The four or six track event is then
analysed in terms of its topology and the topology cut exploits the fact
that the charged tracks originate from four taus in the signal, which
themselves originate from two heavily boosted pseudo-scalars. To avoid
the effects of pile-up, the analysis is heavily track-based with almost no reliance on the calorimeter. 

\begin{figure}[t]
\begin{center}
\includegraphics[width=0.45\textwidth]{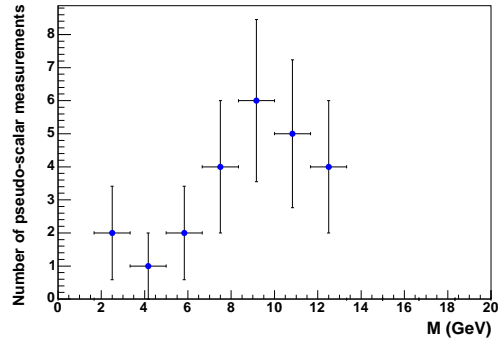}
\caption{A typical $a$ mass measurement.}
\label{fig:amassdata}
\end{center}
\end{figure}

Accurate measurement of the proton energies allows one to constrain
the kinematics of the central system (in particular its invariant mass
and mean rapidity are known). We can also extract the masses of the
$h$ and the $a$ on an event-by-event basis. The mass of the $h$ is
straightforward of course (it is measured directly by the forward
detectors) and a precision below 1 GeV can be obtained with just a
handful of events. The measurement of the pseudo-scalar mass is more
interesting and potentially very important. The proton measurements
fix $p_z$ and $p_{x,y} \approx 0$ for the central system. In addition,
the tau pairs are highly boosted, which means they are collinear with
their parent pseudo-scalars. That means that the four-momentum of each
pseudo-scalar is approximately proportional to the observed (track)
four-momentum. The two unknown constants of proportionality (i.e. the
missing energy fractions) are overconstrained, since we have three
equations from the proton detectors. The result is that we can solve
for the pseudo-scalar masses, with four measurements per
event. Fig.~\ref{fig:amassdata} shows a typical distribution of $a$
masses based on 180~fb$^{-1}$ of data collected at $3 \times 10^{33}$~cm$^{-2}$s$^{-1}$.

\begin{table*}
\caption{The table of cross-sections for the signal and backgrounds. All cross-sections are in femtobarns. The overlap (OLAP) background is computed at a luminosity of $10^{34}$~cm$^{-2}$~s$^{-1}$.}
\begin{tabular}{|c|c|c|c|c|c|c|c|} \hline
 &  \multicolumn{3}{c|}{CEP} & DPE & OLAP &\multicolumn{2}{c|}{QED} \\ \hline
Cut & $H$ & $b\bar{b}$ &$gg$ & $b\bar{b}$& $b\bar{b}$ & $4\tau$ & $2\tau$~$2l$ \\ \hline
$p_{T0}^\mu$, $\xi_1$, $\xi_2$, $M$     & 0.442 & 25.14 & 1.51$\times10^{3}$ & 1.29$\times10^{3}$& 1.74$\times10^{6}$ & 0.014 & 0.467 \\ \hline
$N_{\mathrm{ch}} =$ 4 or 6 & 0.226 & 1.59 & 28.84 & 1.58$\times10^{2}$ & 1.44$\times10^{4}$& 0.003 & 0.056\\ \hline
$Q_h = 0$ & 0.198 & 0.207 & 3.77 & 18.69 & 1.29$\times10^{3}$ & 5$\times10^{-4}$& 0.010 \\ \hline
Topology & 0.143 & 0.036 & 0.432 & 0.209 & 1.84 & - & $<$0.001 \\ \hline
$p_T^{\mu}$, isolation & 0.083 & 0.001 & 0.008 & 0.003 & 0.082 & - & - \\ \hline
$p_T^{1, \not{\mu}}$ & 0.071  & 5$\times$10$^{-4}$ & 0.004 & 0.002 & 0.007  & -& - \\ \hline
$m_a > 2 m_\tau$ & 0.066 & 2$\times$10$^{-4}$ & 0.001 & 0.001 & 0.005 & - & - \\ \hline
\end{tabular}\\[2pt]
\label{tb:sigmas}
\end{table*}

\begin{figure}[t]
\centering
\includegraphics[width=0.4\textwidth]{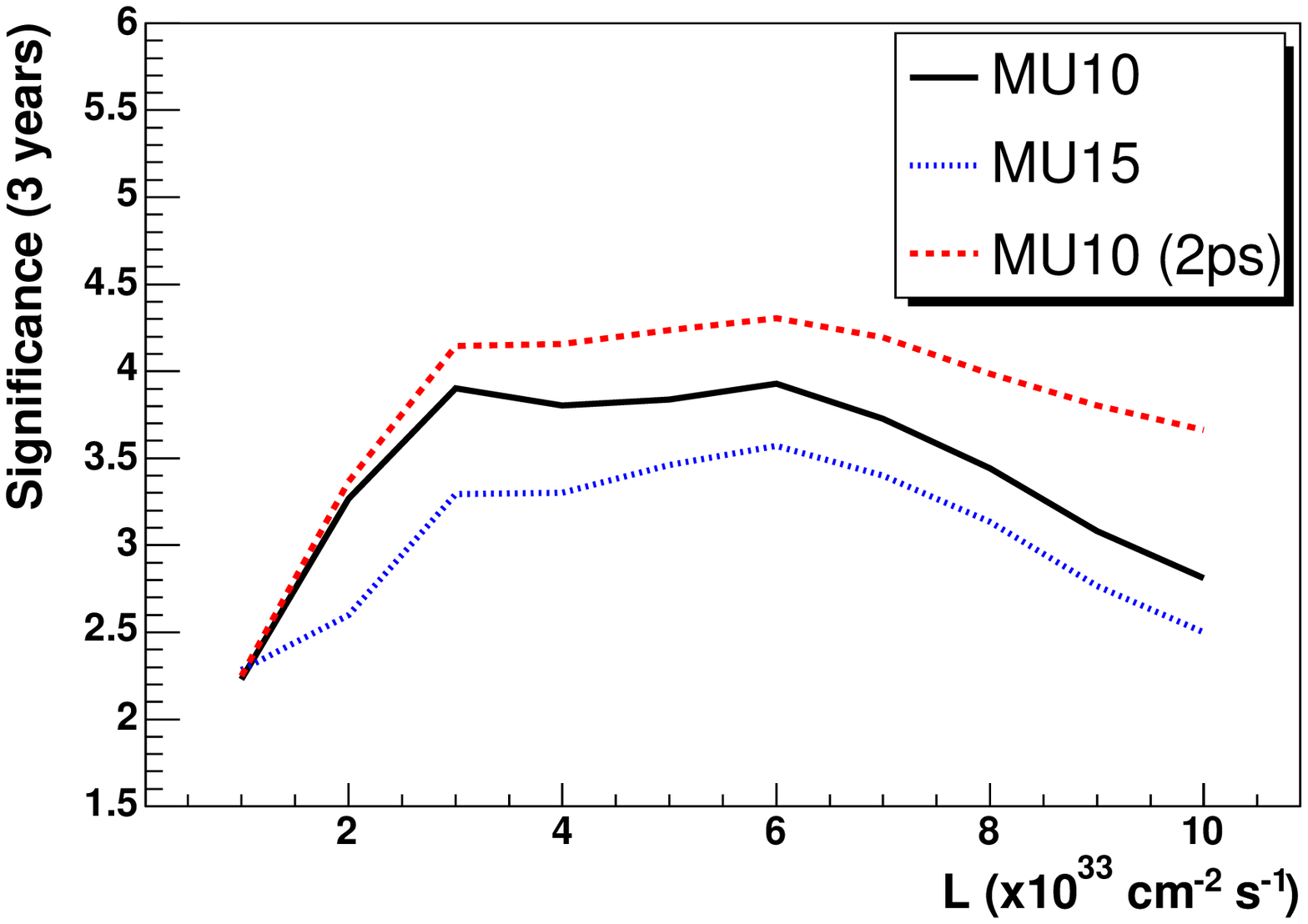}
\includegraphics[width=0.4\textwidth]{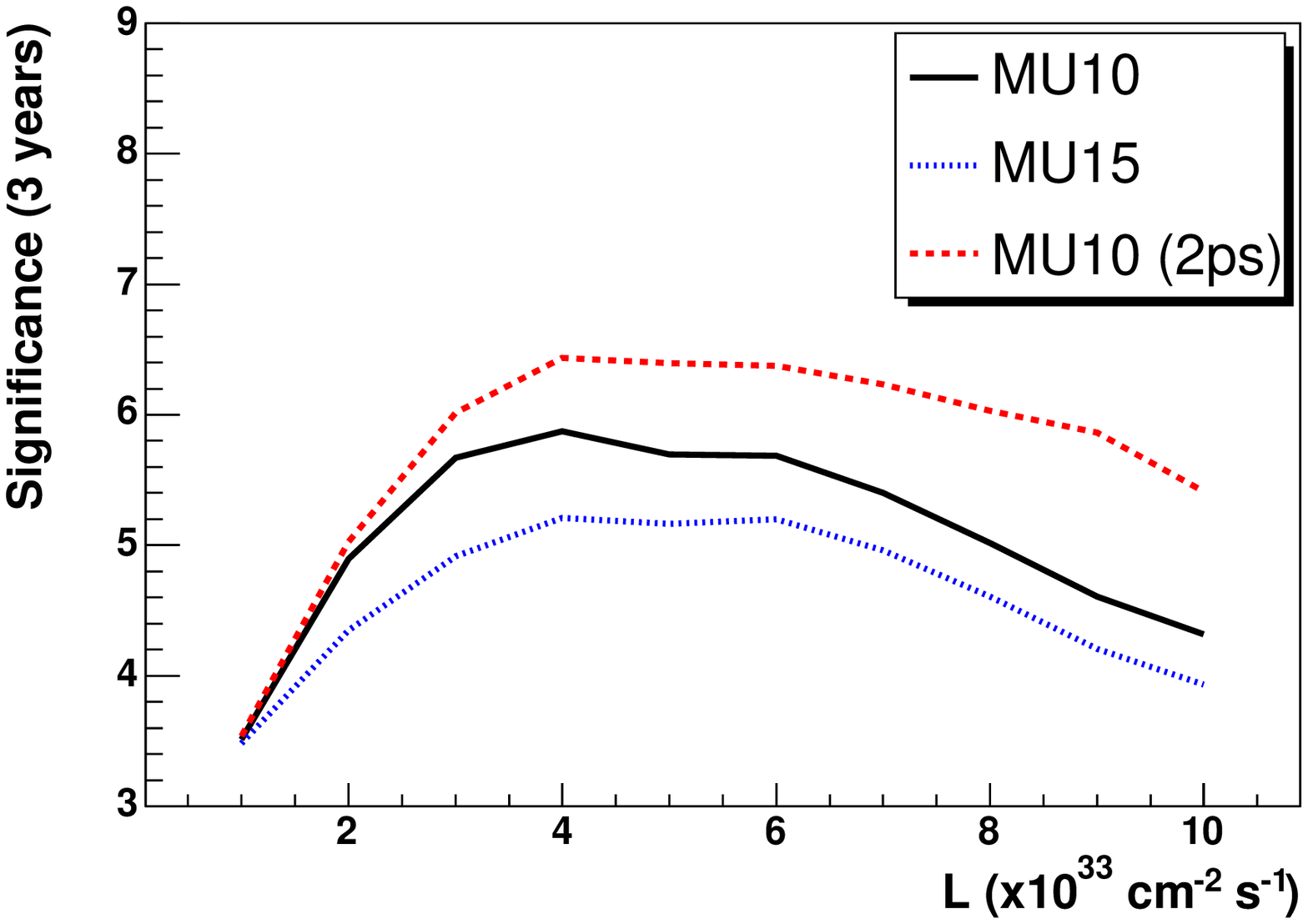}
\caption{Upper: The significance for three years of data acquisition
  at each luminosity. Lower: Same as (a) but with twice the data.}
\label{fig:signif}
\end{figure}

In Table 2 I show the bottom line numbers for three different trigger
scenarios and three different instantaneous luminosities. The key point
is that, although we can expect only a handful of signal events, the
background is under control. Remember, we need only a few events in
order to extract the masses of both the $a$ and the $h$. The
statistical significance of any discovery is estimated in
Fig.~\ref{fig:signif}. The lower plot might pertain if the
signal rate were doubled (recall the theoretical uncertainty permits
it) or if data from ATLAS and CMS were combined or if other
leptonic triggers are used \cite{Albrow:2008pn}.

\section{Concluding remarks}
Central exclusive production is a very attractive prospect for the
LHC. A very broad programme of physics can be pursued for very little
additional expenditure. Measurements from CDF at the Tevatron are
very encouraging and support the validity of the theoretical
calculations: the theory is probably not too far off the
mark. Moreoever, one of the highlights of the past couple of years has
been the demonstration that high luminosity backgrounds can be brought
under control. This talk has focussed only upon Higgs physics and has
placed particular emphasis on the possibility that CEP could be the
only way one could observe at the LHC a Higgs boson that decays predominantly to
four taus: something that could be fairly generic feature
of SUSY models with an enlarged Higgs sector (such as the NMSSM). 

\begin{table}[h]
\caption{Expected number of signal (S) and background (B) events for
  the three trigger scenarios assuming that the data are collected at
  a fixed instantaneous luminosity over  a three year period. We
  assume the integrated luminosity acquired each year is 10~fb$^{-1}$,
  50~fb$^{-1}$ and 100~fb$^{-1}$ at an instantaneous luminosity of
  1$\times10^{33}$~cm$^{-2}$~s$^{-1}$ (row 1),
  5$\times10^{33}$~cm$^{-2}$~s$^{-1}$ (row 2) and 10
  $\times10^{33}$~cm$^{-2}$~s$^{-1}$ (row 3). \label{eventnumbers}}
\begin{tabular}{|c|c|c||c|c||c|c|}
\hline
L & \multicolumn{2}{c|}{MU10} &  \multicolumn{2}{c|}{MU15} &  \multicolumn{2}{c|}{MU10 (2~ps)} \\
   & S & B & S & B & S & B \\
 \hline
 $1$& 1.3 & 0.02 & 1.0 & 0.01 & 1.3 & 0.02\\
 $5$ & 3.7 & 0.14 & 2.9 & 0.08 & 3.7 & 0.07\\
 $10$ & 3.3 & 0.36 & 2.5 & 0.20 & 3.3 & 0.11\\
\hline
\end{tabular}\\[2pt]
\end{table}

\section*{Acknowledgements}
I should like to thank the workshop organizers, both for
inviting me to deliver this talk and for their very generous hospitality.


\begin{thebibliography}{9}

\bibitem{Albrow:2008pn}
  M.~G.~Albrow {\it et al.}  [FP420 R\&D Collaboration],
  ``The FP420 R\&D Project: Higgs and New Physics with forward protons at the
  LHC,''
  arXiv:0806.0302 [hep-ex].

\bibitem{Khoze:2001xm}
  V.~A.~Khoze, A.~D.~Martin and M.~G.~Ryskin,
  Eur.\ Phys.\ J.\  C {\bf 23} (2002) 311
  [arXiv:hep-ph/0111078].

\bibitem{Forshaw:2005qp}
  J.~R.~Forshaw,
  ``Diffractive Higgs production: Theory,''
  arXiv:hep-ph/0508274.

\bibitem{Shuvaev:1999ce}
  A.~G.~Shuvaev, K.~J.~Golec-Biernat, A.~D.~Martin and M.~G.~Ryskin,
  Phys.\ Rev.\  D {\bf 60} (1999) 014015
  [arXiv:hep-ph/9902410].

\bibitem{Martin:2001ms}
  A.~D.~Martin and M.~G.~Ryskin,
  Phys.\ Rev.\  D {\bf 64} (2001) 094017
  [arXiv:hep-ph/0107149].

\bibitem{Martin:2008nx}
  A.~D.~Martin, V.~A.~Khoze and M.~G.~Ryskin,
  ``Rapidity gap survival probability and total cross sections,''
  arXiv:0810.3560 [hep-ph].

\bibitem{Aaltonen:2007hs}
  T.~Aaltonen {\it et al.}  [CDF Collaboration],
  Phys.\ Rev.\  D {\bf 77}, 052004 (2008)
  [arXiv:0712.0604 [hep-ex]].

\bibitem{Monk:2005ji}
  J.~Monk and A.~Pilkington,
  Comput.\ Phys.\ Commun.\  {\bf 175} (2006) 232
  [arXiv:hep-ph/0502077].

\bibitem{Cox:2005if}
  B.~E.~Cox {\it et al.},
  Eur.\ Phys.\ J.\  C {\bf 45} (2006) 401
  [arXiv:hep-ph/0505240].

\bibitem{Heinemeyer:2007tu}
  S.~Heinemeyer, V.~A.~Khoze, M.~G.~Ryskin, W.~J.~Stirling, M.~Tasevsky and G.~Weiglein,
  Eur.\ Phys.\ J.\  C {\bf 53} (2008) 231
  [arXiv:0708.3052 [hep-ph]].

\bibitem{Barate:2003sz}
  R.~Barate {\it et al.}  [LEP Working Group for Higgs boson searches],
  Phys.\ Lett.\  B {\bf 565} (2003) 61
  [arXiv:hep-ex/0306033].

\bibitem{Schael:2006cr}
  S.~Schael {\it et al.}  [ALEPH Collaboration],
  Eur.\ Phys.\ J.\  C {\bf 47} (2006) 547
  [arXiv:hep-ex/0602042].

\bibitem{Cox:2007sw}
  B.~E.~Cox, F.~K.~Loebinger and A.~D.~Pilkington,
  JHEP {\bf 0710} (2007) 090
  [arXiv:0709.3035 [hep-ph]].

\bibitem{Forshaw:2007ra}
  J.~R.~Forshaw, J.~F.~Gunion, L.~Hodgkinson, A.~Papaefstathiou and A.~D.~Pilkington,
  JHEP {\bf 0804} (2008) 090
  [arXiv:0712.3510 [hep-ph]].

\bibitem{Ellwanger:2005dv}
  U.~Ellwanger and C.~Hugonie,
  Comput.\ Phys.\ Commun.\  {\bf 175} (2006) 290
  [arXiv:hep-ph/0508022].

\bibitem{Dermisek:2005ar}
  R.~Dermisek and J.~F.~Gunion,
  Phys.\ Rev.\ Lett.\  {\bf 95} (2005) 041801
  [arXiv:hep-ph/0502105].

\bibitem{Dermisek:2005gg}
  R.~Dermisek and J.~F.~Gunion,
  Phys.\ Rev.\  D {\bf 73} (2006) 111701
  [arXiv:hep-ph/0510322].





\end{thebibliography}
\end{document}